\documentclass{article}
\usepackage{./share/spconf}
\usepackage{./share/package_macros2}
\newacronym{IoT}{IoT}{Internet of Things}
\newacronym{LPWAN}{LPWAN}{Low Power Wide Area Network}
\newacronym{WLAN}{WLAN}{Wireless Local Area Network}
\newacronym{PHY}{PHY}{physical layer}
\newacronym{SDR}{SDR}{software defined radio}
\newacronym{CFO}{CFO}{carrier frequency offset}
\newacronym{SFO}{SFO}{sampling frequency offset}
\newacronym{CSS}{CSS}{chirp spread spectrum}
\newacronym{MAC}{MAC}{Medium Access Control}
\newacronym{CRC}{CRC}{cyclic redundancy check}
\newacronym{DFT}{DFT}{discrete Furrier transform}
\newacronym{BER}{BER}{bit error rate}
\newacronym{SER}{SER}{symbol error rate}
\newacronym{SNR}{SNR}{signal-to-noise ratio}
\newacronym{SoC}{SoC}{system on chip}
\newacronym{TCDM}{TCDM}{tightly coupled data memory}
\newacronym{FFT}{FFT}{fast Furrier transform}
\newacronym{USRP}{USRP}{universal software radio peripheral}
\newcommand{\rdbsec}{\vspace{0mm}}
\newcommand{\rdasec}{\vspace{0mm}}
\newcommand{\rdatext}{\vspace{-3mm}}
\newcommand{\rdextra}{\vspace{-2mm}}

\usepackage{titlesec}
\titlespacing\section{0pt}{10pt plus 4pt minus 2pt}{0pt plus 2pt minus 2pt}
\titlespacing\subsection{0pt}{10pt plus 4pt minus 2pt}{0pt plus 2pt minus 2pt}
\titlespacing\subsubsection{0pt}{10pt plus 4pt minus 2pt}{0pt plus 2pt minus 2pt}
\titleformat{\section}[block]{\bfseries\filcenter}{\thesection.}{1em}{\MakeUppercase}
\titleformat{\subsection}[block]{\bfseries}{\thesubsection.}{1em}{}
\titleformat{\subsubsection}[block]{\em}{\thesubsubsection.}{1em}{}

\title{LoRa Digital Receiver Analysis and Implementation}

\name{\vspace{-5mm} Reza Ghanaatian, Orion Afisiadis, Matthieu Cotting, and Andreas Burg} 

\address{Telecommunication Circuits Laboratory, 
	\'{E}cole polytechnique f\'{e}d\'{e}rale de Lausanne, Switzerland}

\begin{document}
%

\maketitle

\rdbsec
\begin{abstract}

Low power wide area network technologies \mbox{(LPWANs)} are attracting attention because they fulfill the need for long range low power communication for the Internet of Things.
LoRa is one of the proprietary LPWAN physical layer (PHY) technologies, which provides variable data-rate and long range by using chirp spread spectrum modulation.
This paper describes the basic LoRa PHY receiver algorithms and studies their performance. 
The LoRa PHY is first introduced and different demodulation schemes are proposed.
The effect of carrier frequency offset and sampling frequency offset are then modeled and corresponding compensation methods are proposed.
Finally, a software-defined radio implementation for the LoRa transceiver is briefly presented.

\end{abstract}

\begin{keywords}
	LPWAN, IoT, LoRa, carrier frequency offset, sampling frequency offset
\end{keywords}

\rdbsec
\section{Introduction}\label{sec:intro}
\rdasec

\gls{IoT} has sparked a lot of interest. It is foreseen that by 2025 Internet nodes will reside in every day objects, e.g., furniture, packages, etc \cite{li20185g} and \cite{raza2017low}. IoT nodes often target energy autonomy with a battery lifetime of 10+ years, which creates business opportunities for new services in various fields such as home automation, traffic control, environmental monitoring, personal health care, etc.

\gls{LPWAN} technologies have recently emerged to complement existing communication standards. While short-range wireless networks such as Bluetooth, Zigbee, and \glspl{WLAN} are designed to cover short distances with different rates and cellular networks are deployed to provide high rate with global coverage, \glspl{LPWAN} are meant to provide low data rate, wide area coverage and high energy efficiency \cite{raza2017low} for the \gls{IoT}.
Several \gls{LPWAN} technologies, such as different generations of 3GPP standards, e.g., NB-IoT, as well as proprietary ultra-low-rate standards, e.g., LoRa and SigFox, are predicted to co-exist in the future to provide connectivity for billions of nodes. 

In this paper we focus specifically on the LoRa technology. LoRa is a proprietary  \gls{PHY} standard, which was developed by Cycleo and acquired by Semtech in 2012~\cite{semtech2015120022}. LoRaWAN$^\text{TM}$ is an open standard proposed by the LoRa$^\text{TM}$ Alliance \cite{sornin2015lorawan} that defines the network architecture and layers above the LoRa \gls{PHY}.
The work of~\cite{centenaro2016long} and \cite{sinha2017survey} provides a high-level system architecture overview of LoRa. Furthermore, several attempts were recently made to systematically explain the LoRa \gls{PHY} properties. In \cite{robyns2018multi} a description of the reverse engineered proprietary LoRa \gls{PHY} is presented and a software decoder using the GNU radio framework is provided. The work of \cite{knight2016decoding} details the modulation and encoding elements that comprise the LoRa PHY. Although many aspects of the LoRa \gls{PHY} are known, an in-depth analysis and detailed algorithmic description of a LoRa receiver is so far missing in the literature.

In this paper, a detailed analysis of the LoRa \gls{PHY} is provided by studying multiple aspects of a LoRa receiver. In particular, we present different structures for LoRa demodulation, we study the synchronization process in LoRa receivers  and we analyze the effect of \gls{CFO} and \gls{SFO} on the receiver performance. Finally, a \gls{SDR} implementation for LoRa is briefly shown.


\rdbsec
\section{LoRa \gls{PHY} Overview}\label{sec:loraphy}
\rdasec
LoRa employs \gls{CSS} modulation, which provides variable data rates by changing the \emph{spreading factor}.
Therefore, this modulation allows to trade throughput for coverage and/or energy consumption \cite{centenaro2016long}.

In this section, we first present the frame structure of LoRa, we then provide a high-level overview of the LoRa \gls{PHY} and finally focus on the modulation and demodulation as the most important blocks of a LoRa transmitter and receiver.

\rdbsec
\subsection{LoRa Frame Structure}
\rdasec
The LoRa \gls{PHY} frame structure is defined in \cite{sornin2015lorawan} and illustrated in Fig.~\ref{fig:loraphy}.
A frame is composed of a preamble with a number of preamble $N_\text{pre}$ upchirps and $4.25$ LoRa symbols as frame delimiters for synchronization, a \gls{PHY}-header containing the frame information, a variable-length \gls{PHY}-payload, and a \gls{CRC}. The \gls{PHY}-header and the \gls{CRC} are optional.

\rdbsec
\subsection{LoRa \gls{PHY} Block Diagram}
\rdasec
Fig.~\ref{fig:loraphy} shows the block diagram of a LoRa transceiver. On the transmitter side, the input bits are first encoded using a Hamming code. Then whitening, interleaving, and Gray indexing are applied before modulation.
LoRa uses \gls{CSS} modulation for the preamble and the data. The LoRa \gls{CSS} modulation is explained in more detail in the next subsection.
The receiver performs synchronization and frequency-offset estimation and compensation prior to demodulation.
Gray indexing, de-interleaving, de-whitening, and Hamming decoding are carried out to recover the information.

\rdbsec
\subsection{LoRa Modulation and Demodulation}
\rdasec

\subsubsection{LoRa Modulation}
\rdasec
\vspace{1mm}

A LoRa \gls{CSS} modulated symbol with spreading factor $\text{SF} \in \{6,7,\cdots,12\}$ is defined as
\begin{IEEEeqnarray}{lll} \label{Eq:LoRaMod} 
x_S(t) = 
\begin{cases}
e^{j2\pi \big(\frac{BW}{2 T_s} t^2 + (\foff - \frac{BW}{2})t \big)}, & 0 \leq t<t_{\text{fold}},\\
e^{j2\pi \big(\frac{BW}{2 T_s} t^2 + (\foff - \frac{3BW}{2})t \big)},& t_{\text{fold}} \leq t<T_{s}, 
\end{cases}
\end{IEEEeqnarray}
where \mbox{$BW \in\{125, 250, 500\}$\,kHz} is the bandwidth, \mbox{$T_s={{2^\text{SF}} \over {BW}}$} is the symbol duration, $\foff$ is the initial frequency of a chirp, which depends on the data symbol $S \in \{0,1,\dots,2^\text{SF}-1\}$ and is defined as $\foff= S  \cdot {BW \over 2^\text{SF}}$, and $t_{\text{fold}}={{2^\text{SF}-S} \over {BW}}$.
LoRa is a spread spectrum technology, which indicates that $2^\text{SF}$ samples are transmitted per LoRa symbol to convey SF bits. To encode the data, the two-sided baseband bandwidth is split into $2^\text{SF}$ frequency steps. The symbol frequency starts at $\foff - {BW \over 2}$ and increases linearly with time until it reaches the Nyquist frequency $BW \over 2$ at $t=t_\text{fold}$, where a \emph{frequency fold} to $- {BW \over 2}$ occurs. Setting $S=0$, results in an \emph{upchirp}, whose frequency continuously increases during the symbol duration.

The discrete-time equation for a LoRa symbol is derived by replacing $t = {n \over f_s}$ in \eqref{Eq:LoRaMod}, where $n$ is the sample index and $f_s$ is the sampling frequency, as
\begin{IEEEeqnarray}{lll} \label{Eq:LoRaMod_discr} 
	x_S[n] = 
	\begin{cases}
		e^{j2\pi \big(\frac{1}{2 \cdot 2^\text{SF}} (\frac{BW}{f_s})^2 n^2 + ({S \over 2^\text{SF}}- {1 \over 2})({BW \over f_s}) n \big)},  \; n \in N_1,\\
		e^{j2\pi \big(\frac{1}{2 \cdot 2^\text{SF}} (\frac{BW}{f_s})^2 n^2 + ({S \over 2^\text{SF}}- {3 \over 2})({BW \over f_s}) n \big)},  \; n \in N_2, 
	\end{cases}
\end{IEEEeqnarray}
where $N_1= \{0,..., n_{\text{fold}}-1\}$, $N_2=\{n_{\text{fold}},...,2^\text{SF}-1\}$, and $n_{\text{fold}}=t_{\text{fold}}f_s$. By setting $f_s = BW$, \eqref{Eq:LoRaMod_discr} is simplified to 
\begin{IEEEeqnarray}{lll} \label{Eq:LoRaMod_discr2} 
x_S[n]=e^{j2\pi \big( {n^2 \over 2 \cdot 2^\text{SF}} + ({S \over 2^\text{SF}}- {1 \over 2}) n \big)}, n \in \{0,1,\cdots,2^\text{SF}-1 \}.
\end{IEEEeqnarray}

\subsubsection{LoRa Demodulation}
\vspace{1mm}	
The received signal is given by
\begin{IEEEeqnarray}{lll} \label{Eq:LoRarx}
y_S[n]= h \, x_S[n]+z[n],
\end{IEEEeqnarray}
where $h$ denotes the block-fading channel, and $z[n]$ is the zero-mean white Gaussian complex-valued noise with variance $\sigma^2$.
A non-coherent demodulator applies $2^\text{SF}$ matched filters with the candidate reference symbols 
\begin{IEEEeqnarray}{lll} \label{eq:corr_demod}
X_k = \sum_{n=0}^{2^\text{SF}-1} y_S[n] x_k\hermis[n] = \sum_{n = 0}^{2^{\text{SF}}-1} e^{j 2 \pi \left(n \left( \frac{S-k}{2^{\text{SF}}} \right) \right)},
\end{IEEEeqnarray}
and retrieves the maximum-likelihood symbol estimate $\hat{S}$ with 
\begin{IEEEeqnarray}{lll} \label{eq:retrieved_symbol}
\hat S = \underset{k}{\mathrm{argmax}} \big( |X_k| \big).
\end{IEEEeqnarray}

Unfortunately, the complexity of the above scheme is high due to the required $2^\text{SF}$ convolutions in \eqref{eq:corr_demod}.
Another, less complex formulation of the LoRa demodulator first multiplies the received symbol with the complex-conjugate of an upchirp, i.e., \emph{dechirping}, as $y_{\text{dc}_S}[n]=y_S[n] x_0\hermis[n]$ followed by a \gls{DFT} of the dechirped signal.
In this way, \eqref{eq:corr_demod} can be written as
\begin{IEEEeqnarray}{lll} \label{eq:DFT_demod}
	[X_1, X_2,\dots, X_{2^\text{SF}-1}] = \text{DFT}(y_\text{dc}[n]),
\end{IEEEeqnarray}
which is fully equivalent, but less complex to compute.
We consider \eqref{eq:DFT_demod} for the analysis provided in Section~\ref{sec:lorarx}.

\begin{figure}[t]
	\centering
	\includegraphics[width=0.5\textwidth]{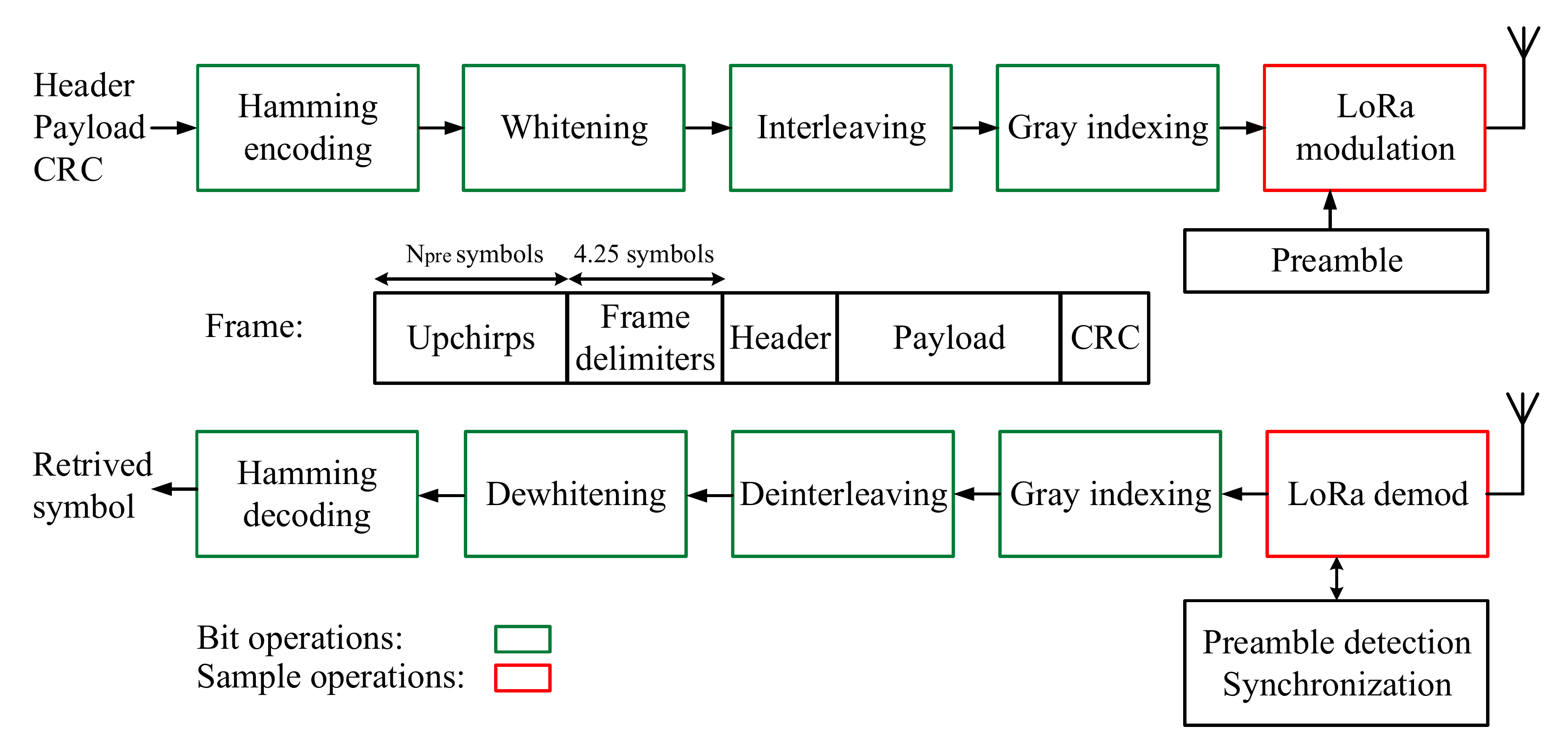}
	\vspace{-8mm}
	\caption{LoRa frame structure \cite{sornin2015lorawan} and \gls{PHY} block diagram.}
	\vspace{-3mm}
	\label{fig:loraphy}
\end{figure}

\rdbsec
\section{LoRa Receiver Analysis}\label{sec:lorarx}
\rdasec
In the previous section, we assumed perfect synchronization and no frequency offset during the LoRa demodulation.
In this section, we discuss the necessary initial synchronization and the receiver in the presence of \gls{CFO} and \gls{SFO}. Specifically, we propose a synchronization algorithm, we study the effect of \gls{CFO} on the synchronization and demodulation, and we provide an analysis on the effect of \gls{SFO} on error-rate performance.

\rdbsec
\subsection{Preamble Detection and Synchronization}
\rdasec
The first step in the receiver is to detect the preamble and to synchronize to the frame boundary. To this end, we exploit the repeating upchirps in the preamble.
More specifically, the synchronization is done in two steps: The receiver collects a block of samples\footnote{We note that by scanning the received signal power level, the receiver has a mean to detect a transmission, while the start of the preamble still needs to be detected.}
and computes \eqref{eq:DFT_demod}.
If at least one of the frequency bins has a magnitude that exceeds a given threshold, the index of the bin with the largest magnitude is noted as $\hat S_\text{pre}$ and the process is continued. If a LoRa preamble is present, this value remains the same during the next DFT blocks and $N_\text{pre}-1$ equal indices $\hat S_\text{pre}$ are detected.
Since the preamble consists of consecutive upchirps, the index $\hat S_\text{pre}$ indicates the time offset in samples between the start of the block collected for synchronization and the start of the received preamble symbol.
Therefore, the receiver synchronizes to the start of the header by skipping $2^\text{SF} - \hat S_\text{pre}$ samples and the initial frame delimiters.

\rdbsec
\subsection{\gls{CFO} Formulation and Robustness Analysis}

Low-cost crystal oscillators have an inherent mismatch with their nominal frequency value and therefore the down-conversion is performed with a different frequency than the up-conversion. In this subsection, we analyze the effect of the resulting \gls{CFO} on the receiver performance.

We denote with $ f_{c_1} $ the carrier frequency that is used for up-conversion, and with $ f_{c_2} $ the frequency that is used for down-conversion. The difference \mbox{$\Delta f_{c} = f_{c_1} - f_{c_2}$} is the carrier frequency offset. For the simplicity of notation we consider only the signal without the noise.
A LoRa symbol after up- and down-conversion and dechirping can be written as
\begin{IEEEeqnarray}{lll}
\label{eq:dechirped_symbol_withOffset}
\tilde{y}_{\text{dc}_S}[n] = \tilde{y}_S[n] x_0\hermis[n] = e^{j2\pi n \big(\frac{S}{2^{\text{SF}}} - \frac{\Delta f_{c}}{f_{s}} \big)},
\end{IEEEeqnarray}
where $\tilde{y}_S[n] = y_S[n] \cdot e^{j2\pi n \frac{\Delta f_{c}}{f_{s}}}$ is the LoRa symbol with \gls{CFO}.
The CFO results in a frequency shift, which can introduce an error in the decision of \eqref{eq:retrieved_symbol}.
More specifically, if the offset is sufficiently large to displace the peak in the Fourier domain by more than half a bin, i.e., $ \frac{\vert\Delta f_{c}\vert}{f_{s}} >  \frac{1}{2\cdot 2^{\text{SF}}}$, a demodulation error will occur even without noise. This offset stays constant during the frame if the \gls{CFO} stays constant over time.



The LoRa demodulator described in the previous section is, however, partially robust against a \gls{CFO} since the \gls{CFO} leads to a time offset in the synchronization equal to $\round{\frac{\Delta f_{c}}{f_{s}} \cdot2^{SF}}$ samples, where $\round \cdot$ denotes the rounding operation. This time offset partially mitigates the \gls{CFO} introduced in \eqref{eq:dechirped_symbol_withOffset}.
However, a residual \gls{CFO} remains, which can still result in a significant error-rate performance degradation. This offset needs to be compensated to prevent a symbol lying in between two frequency bins, which increases the sensitivity to noise.

\begin{figure}
	\centering
	\begin{tikzpicture}	
	\pgfplotsset{grid style={dotted}}
    \begin{semilogyaxis}[
        width = 0.97\columnwidth,
        height = 0.75\columnwidth,
        xlabel = {$\text{E}_\text{b} / \text{N}_\text{0}$ (dB)},
        xlabel style={yshift=0.65em},
        ylabel = {BER},
        xmin = -4, xmax = 8, 
        ymin = 1e-4, ymax = 6e-1, 
        grid = both,
        legend style={legend pos=south west,font=\tiny},
        legend cell align=left,
        legend entries={No offset, offset = $5$\,KHz, offset = $10$\,KHz},
        name=plot1,
        ]
       	\addlegendimage{black, thick, solid};
       	\addlegendentry{No offset};
       	
     	\addlegendimage{brown, thick, solid, mark=*};
     	\addlegendentry{ $10$\,kHz};
     	       	
     	\addlegendimage{cyan, thick, solid, mark=pentagon*};
     	\addlegendentry{$10$\,kHz; \;\;\; Time-offset};
       	
       	\addlegendimage{green, thick, solid, mark=triangle*};
       	\addlegendentry{$10.1$\,kHz; Time-offset};
       	
       	\addlegendimage{blue,thick, solid, mark=square*};
       	\addlegendentry{$10$\,kHz; \;\;\; Time-offset and compensation};
       	
		\addlegendimage{red,thick, solid, mark=x};
		\addlegendentry{ $10.1$\,kHz; Time-offset and compensation};
       	

	\addplot[black, thick, solid] table[x index=0,y index=1]{./figs/data/BER_completeSystem_CFO0k.dat};
	
	\addplot[brown, thick, solid, mark=*] table[x index=0,y index=1]{./figs/data/BER_completeSystem_CFO10k_noalign_nocomp.dat};
	\addplot[cyan, thick, solid, mark=pentagon*] table[x index=0,y index=1]{./figs/data/BER_completeSystem_CFO10k_align_nocomp.dat};
	\addplot[blue, thick, solid, mark=square*,] table[x index=0,y index=1]{./figs/data/BER_completeSystem_CFO10k_align_comp.dat};
	
	\addplot[green, thick, solid, mark=triangle*] table[x index=0,y index=1]{./figs/data/BER_completeSystem_CFO10.1k_align_nocomp.dat}; 
	\addplot[red, thick, solid, mark=x,] table[x index=0,y index=1]{./figs/data/BER_completeSystem_CFO10.1k_align_comp.dat};

    \end{semilogyaxis}
\end{tikzpicture}
	\vspace{-4mm}
	\caption{BER for \gls{CFO} of $10$\,kHz and $10.1$\,kHz.}
	\label{fig:ber_cfo_mis-synch}
	\vspace{-3mm}
\end{figure}
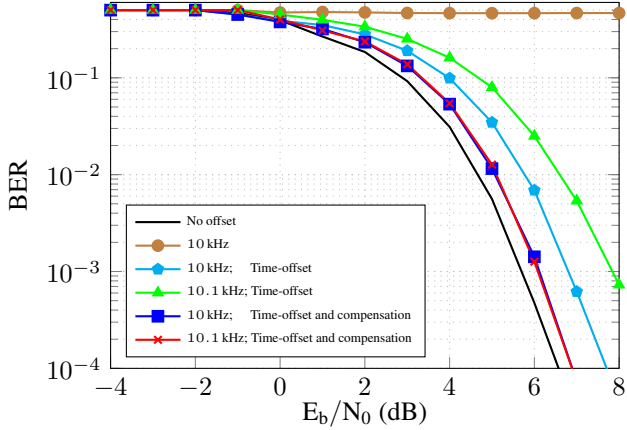

%

The \gls{CFO} results in a phase offset \mbox{$\Delta{\phi} = 2 \pi {\Delta f_{c} \over f_{s}} \cdot 2^{\text{SF}}$} between two samples with the same index in consecutive upchirps. The residual part of this offset can be estimated by taking the average across the entire symbol as \rdatext
\begin{IEEEeqnarray}{lll}
\label{eq:deltaphi}
\hat{\Delta \phi} = \arg \Bigg( \sum_{n=0}^{2^\text{SF}-1}\tilde{y}_S[n] \cdot \tilde{y}\hermis_S[n+2^\text{SF}] \Bigg).
\end{IEEEeqnarray}
We note that the estimation becomes more accurate by taking the average among all the upchirps in the preamble.
The residual \gls{CFO} can be compensated as
\begin{IEEEeqnarray}{lll}
\label{eq:cfo_comp}
\hat{y}_S[n] = \tilde{y}_S[n] \cdot e^{jn {\hat{\Delta \phi} \over 2^\text{SF}}}.
\end{IEEEeqnarray}

To study the effect of a \gls{CFO} and the proposed compensation method, we use Monte-Carlo simulations, where we generate a LoRa signal and add the \gls{CFO} according to \eqref{eq:dechirped_symbol_withOffset}.
Fig.~\ref{fig:ber_cfo_mis-synch} shows the \gls{BER} of the system illustrated in Fig.~\ref{fig:loraphy} with SF $=8$ and Hamming (4, 8). The offset values chosen in the simulation are $10$\,kHz and $10.1$\,kHz, which both correspond approximately to $10$ ppm.
As can be seen, the synchronization with the time-offset can improve the performance for both \gls{CFO} values, while the residual \gls{CFO} is different for them.
This residual \gls{CFO} is then compensated by using the algorithm as in \eqref{eq:cfo_comp}, where the estimation is performed using the entire preamble.
We note that the small remaining difference in comparison to the ideal system without \gls{CFO} is due to the small remaining inter-symbol interference (ISI) because of the synchronization with a time-offset.

\rdbsec
\vspace{-3mm}
\subsection{\gls{SFO} Formulation and Performance Analysis}


The mismatch between the transmitter and receiver oscillators also results in different sampling frequencies,
and therefore, the sampled LoRa signal at the receiver experiences a sampling frequency offset. In this subsection, we study the effect of this offset on the receiver performance.

The received LoRa symbols undergo low-pass filtering with bandwidth $BW$ and 
are sampled with the receiver sampling frequency $f'_s \neq BW$. We consider again the signal without noise.
To describe the impact of the \gls{SFO} throughout the entire frame, we introduce the LoRa symbol index $d\in\{0,1,\dots\}$ and write the ISI free samples of the $d$-th symbol as \rdatext
\begin{IEEEeqnarray}{lll} \label{Eq:YSFO} 
	y^d_{S}[n] = 
	\begin{cases}
		e^{j2\pi \big(\frac{BW}{2 T_s} ({n \over f'_s}+\Delta T_d)^2 + (S \cdot {BW \over 2^\text{SF}} - \frac{BW}{2})({n \over f'_s}+\Delta T_d) \big)},  \\
		e^{j2\pi \big(\frac{BW}{2 T_s} ({n \over f'_s}+\Delta T_d)^2 + (S \cdot {BW \over 2^\text{SF}} - \frac{3BW}{2})({n \over f'_s}+\Delta T_d) \big)},  \\
	\end{cases} 
\end{IEEEeqnarray}
where $\Delta T_d=d\left(\frac{2^{\text{SF}}}{f'_s}-T_s\right)$ is the time offset of the $d$-th symbol, $\lfloor -\Delta T_d f'_s \rfloor < n < \lceil (T_s-\Delta T_d )f'_s \rceil$, and the conditions of each equation are pruned for notational simplicity.

Even if we ignore the ISI, ideally, the signal needs to be resampled to be able to be demodulated as in \eqref{eq:corr_demod} and \eqref{eq:retrieved_symbol} due to the sampling rate mismatch between $f'_s$ and $BW$.
However, resampling is a complex operation. Instead, we only assume that the receiver is able to generate a reference upchirp that matches the bandwidth of the transmitter upchirp as \mbox{$y'_\text{ref}[n] = e^{j2\pi \big({BW \over 2T_s} ({n \over f'_s})^2 -{BW \over 2}({n \over f'_s}) \big)}$}.
The dechirped signal for $d$-th LoRa symbol is then derived by multiplying the received signal	$y^d_{S}[n]$ with the complex-conjugate of $y'_\text{ref}[n]$. After some simplifications and removing the constant phase offsets we obtain \rdatext
\begin{IEEEeqnarray}{lll} \label{Eq:YnSFO_dc} 
	y^d_{\text{dc}_S}[n] = 
	\begin{cases}
		e^{j2\pi n \big( {S \over 2^\text{SF}} (\bwf) + d (\bwfsq - \bwf) \big)},   \\
		e^{j2\pi n \big( ({S \over 2^\text{SF}}-1) (\bwf) + d (\bwfsq - \bwf)\big)}. 
	\end{cases}
\end{IEEEeqnarray}
Subsequently, the \gls{DFT} of the dechirped signal is computed as \rdatext \rdextra
\begin{IEEEeqnarray}{lll}\label{Eq:YSFO_DFT}  
	X^d_k & = \text{DFT}(y^d_{\text{dc}_S}[n]) = \sum_{n=0}^{2^\text{SF}-1} y^d_{\text{dc}_S}[n] \cdot e^{-j2\pi n {k \over 2^\text{SF}}} \nonumber
	\\
	& = \sum_{n \in N^d_{1}} e^{j{2\pi n \over 2^\text{SF}} \big( S \bwf +2^\text{SF}d (\bwfsq - \bwf) -k \big)} \\	\nonumber
	& + \sum_{n \in N^d_{2}} e^{j{2\pi n \over 2^\text{SF}} \big( (S-2^\text{SF}) \bwf +2^\text{SF}d (\bwfsq - \bwf) -k \big)},
\end{IEEEeqnarray}
where $N^d_{1}$ and $N^d_{2}$ are the sets of ISI free indices before and after folding for $d$-th symbol, respectively. For a small frequency offset such that $e^{j2\pi \frac{BW}{f'_s}}\approx 1$, the above equation is simplified to \rdatext
\begin{IEEEeqnarray}{lll}\label{Eq:SFO_DFT2} 
	X^d_{k} \approx \sum_{n=0}^{2^\text{SF}-1} e^{j{2\pi n \over 2^\text{SF}}\big( \big[S \bwf +2^\text{SF}d (\bwfsq - \bwf)\big] -k \big)}.
\end{IEEEeqnarray}
We can observe the effect of the \gls{SFO} by considering the term $\bwf$ in \eqref{Eq:SFO_DFT2}. We first consider the case of no offset, i.e., \mbox{$\bwf=1$}. In such a case, $X^d_k$ will be maximized and equal to $2^\text{SF}$ for $k=S$ and zero elsewhere.
However, for the case of an offset, we observe a re-scaling of the symbol location to ${2\pi} S \bwf$ with respect to the receiver frequency axis (i.e., a symbol-dependent offset of $S(\bwf-1)$) and a frequency-offset ${2\pi} d (\bwfsq - \bwf)$ that depends only on the index of the received symbol $d$. The effect of the re-scaling of the frequency axis is negligible and does not change throughout the received frame, but the frequency-offset term leads to \mbox{\emph{side-lobes}} in the frequency bins other than the desired bin. These sidelobes render the decision in \eqref{eq:retrieved_symbol} suboptimal and lead to an increase in the noise sensitivity. The mismatch between $f'_s$ and $BW$ causes a drift of the sidelobes as $d$ increases and will eventually lead to a drift of the peak in the Fourier domain into the adjacent symbol to $S$ which results in a constant demodulation error, i.e., an error floor. Specifically, assuming that $f'_s > BW$ with no loss of generality, the sample drift occurs at the first sample $n$ in $d$-th symbol that satisfies ${n+1+d2^\text{SF} \over f'_s} < {n+d2^\text{SF} \over BW}$. Therefore, the frames with number of symbols larger than $d$ experience such an error floor.

Motivated by the above, we propose to discard a sample whenever half of a sample has drifted into the adjacent symbol. More specifically, the receiver can find the indices  $n$ and $d$ that satisfy ${n+1/2+d2^\text{SF} \over f'_s} < {n+d2^\text{SF} \over BW}$ and discard sample $n$ from $d$-th symbol. 
This measure \emph{re-aligns} the symbol boundaries and prevents the accumulation of error due to a \gls{SFO}. We note that the re-alignment resolution is half of a sample duration, i.e., $1 \over 2 f'_s$. Oversampling can improve this resolution and thus lower the error rate due to the sample drift from the adjacent symbol.

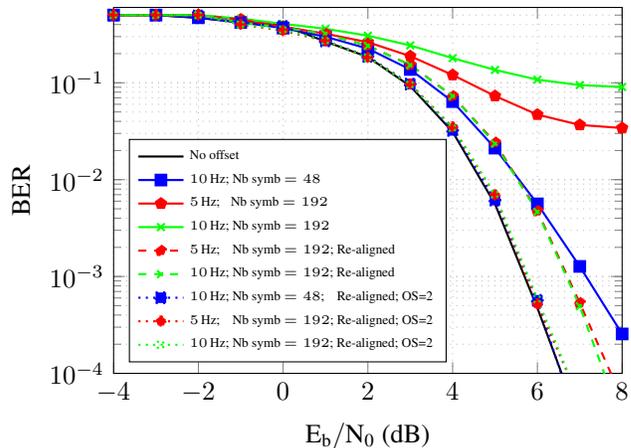
\begin{figure}[!t]
	\centering
	\begin{tikzpicture}
	\pgfplotsset{grid style={dotted}}
	\begin{semilogyaxis}[
		width = 0.97\columnwidth,
		height = 0.75\columnwidth,
		xlabel = {$\text{E}_\text{b} / \text{N}_\text{0}$ (dB)},
		ylabel = {BER},
		xmin = -4, xmax = 8, 
		ymin = 1e-4, ymax = 6e-1, 
		grid = both,
		legend style={legend pos=south west,font=\tiny},
		legend cell align=left,
		legend entries={No offset, offset = $5$\,KHz, offset = $10$\,KHz},
		name=plot2,
		]
		\addlegendimage{black, thick, solid};
		\addlegendentry{No offset};
		
		
		\addlegendimage{blue, thick, mark=square*}; 
		\addlegendentry{$10$\,Hz; Nb symb $= 48$};
		
		\addlegendimage{red, thick, mark=pentagon*};
		\addlegendentry{$5$\,Hz;\;\; Nb symb $= 192$};
		
		\addlegendimage{green, thick, mark=x};
		\addlegendentry{$10$\,Hz; Nb symb $= 192$};
		
		\addlegendimage{red, thick, , dashed, mark=pentagon*};
		\addlegendentry{$5$\,Hz;  \;\, Nb symb $= 192$; Re-aligned};
		
		\addlegendimage{green, thick, dashed, mark=x};
		\addlegendentry{$10$\,Hz; Nb symb $= 192$; Re-aligned};

		\addlegendimage{blue, thick, dotted, mark=square*}; 
		\addlegendentry{$10$\,Hz; Nb symb $= 48$; \;\, Re-aligned; OS=2};
		
		\addlegendimage{red, thick, dotted, mark=pentagon*};
		\addlegendentry{$5$\,Hz; \;\, Nb symb $= 192$; Re-aligned; OS=2};
		
		\addlegendimage{green, thick, dotted, mark=x};
		\addlegendentry{$10$\,Hz; Nb symb $= 192$; Re-aligned; OS=2};


		\addplot[black, thick, solid] table[x index=0,y index=1]{./figs/data/BERmoresnr_offset0_nbsym48.dat};			
		
		\addplot[blue, thick, solid, mark=square*] table[x index=0,y index=1]{./figs/data/BERmoresnr_offset10_nbsym48.dat};
	
		\addplot[red, thick, solid, mark=pentagon*,] table[x index=0,y index=1]{./figs/data/BERmoresnr_offset5_nbsym192.dat};
		\addplot[green, thick, solid, mark=x,] table[x index=0,y index=1]{./figs/data/BERmoresnr_offset10_nbsym192.dat};
		
		\addplot[red, thick, dashed, mark=pentagon*, ] table[x index=0,y index=1]{./figs/data/BERmoresnr_comp_offset5_nbsym192.dat};
		\addplot[green, thick, dashed, mark=x, ] table[x index=0,y index=1]{./figs/data/BERmoresnr_comp_offset10_nbsym192.dat};
		
		\addplot[blue, thick, dotted, mark=square*] table[x index=0,y index=1]{./figs/data/BERmoresnr_comp_offset10_nbsym48_os2.dat};
		\addplot[red, thick, dotted, mark=pentagon*, ] table[x index=0,y index=1]{./figs/data/BERmoresnr_comp_offset5_nbsym192_os2.dat};
		\addplot[green, thick, dotted, mark=x, ] table[x index=0,y index=1]{./figs/data/BERmoresnr_comp_offset10_nbsym192_os2.dat};
		
	\end{semilogyaxis}
\end{tikzpicture}
	\vspace{-4mm}
	\caption{BER  for \gls{SFO} of $5$\,Hz and $10$\,Hz.}
	\label{fig:SFO}
	\vspace{-4mm}
\end{figure}

To study the effect of a \gls{SFO} and the proposed compensation method, we use Monte-Carlo simulation, 
where a LoRa signal is generated with bandwidth $BW$ at the transmitter and is sampled with frequency $f'_s$ at the receiver.
Fig.~\ref{fig:SFO} demonstrates the \gls{BER} of the system illustrated in Fig.~\ref{fig:loraphy} with \mbox{SF $=8$} and Hamming (4, 8).
The offset values chosen in the simulation are $5$\,Hz and $10$\,Hz, which correspond to $20$ and $40$ ppm, respectively.
As can be seen, the performance degradation becomes more severe by increasing the offset value as well as the number of symbols in the frame. However, re-aligning the symbol boundaries according to the proposed method prevents the error-floor in the frame with large number of symbols. 
Furthermore, oversampling with only a factor of $2$ results in a sufficient re-alignment resolution that almost entirely compensates the performance degradation.


\rdbsec
\rdextra
\section{\gls{SDR} Implementation}\label{sec:implementation}

The LoRa transceiver was implemented using National Instrument (NI) \glspl{USRP} and was tested with the commercial HOPERF RFM95 LoRa radio chip.
We have used \mbox{NI-USRP 2920} devices to reverse-engineer and understand the LoRa PHY according to the block diagram in Fig.~\ref{fig:loraphy}. To this end, we receive and analyze different transmitted messages from a LoRa radio that have special structures and can help to extract the parameters of the different blocks.
The reversed-engineered LoRa transmitter and receiver are finally verified by decoding the LoRa frames transmitted by the commercial radio, as shown in Fig.~\ref{fig:GUI}.

{\begin{figure}
	\centering
    \subfloat{\includegraphics[width=0.18\textwidth]{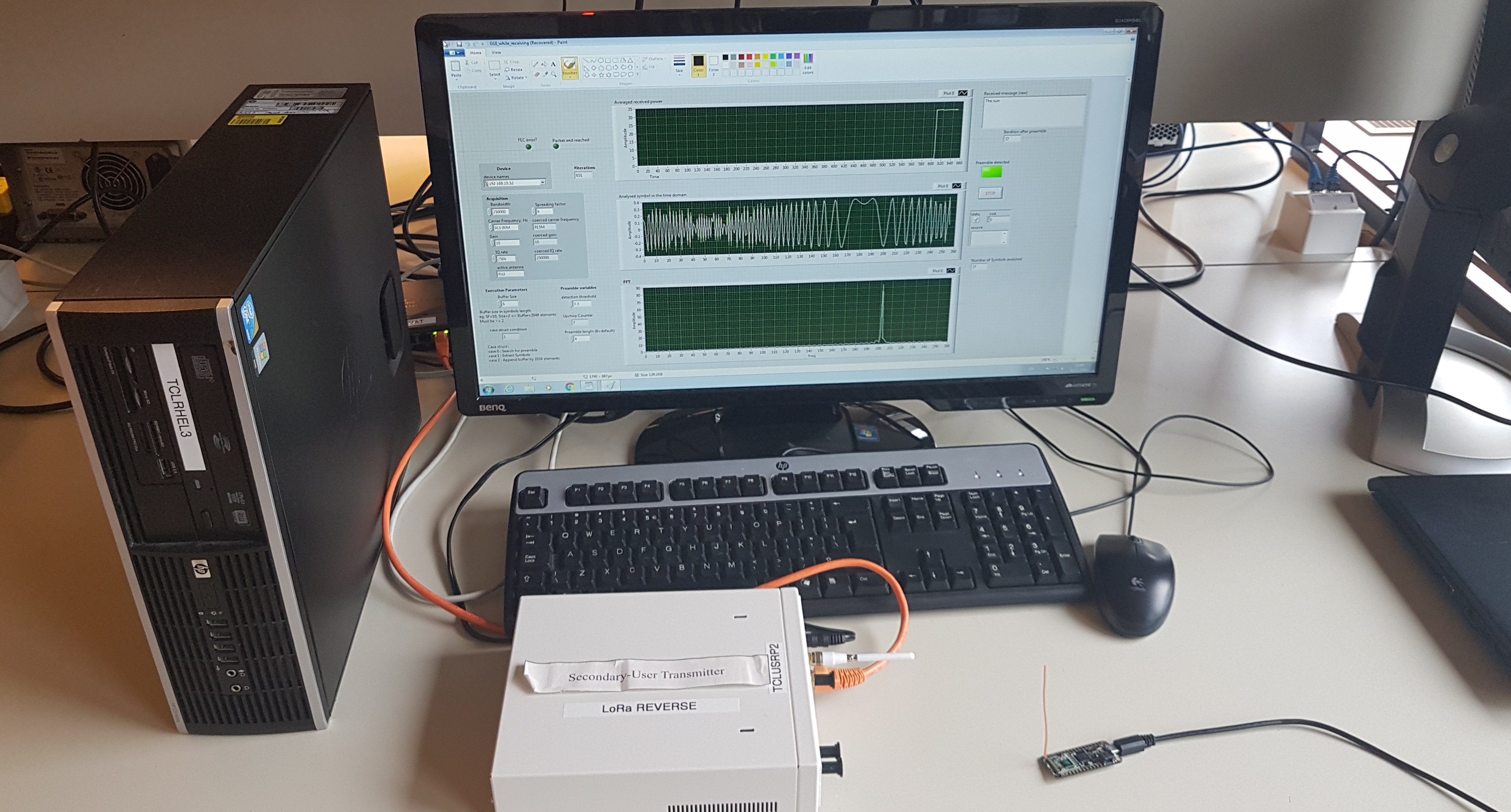}} \hspace{4mm}
	\subfloat{\includegraphics[width=0.18\textwidth]{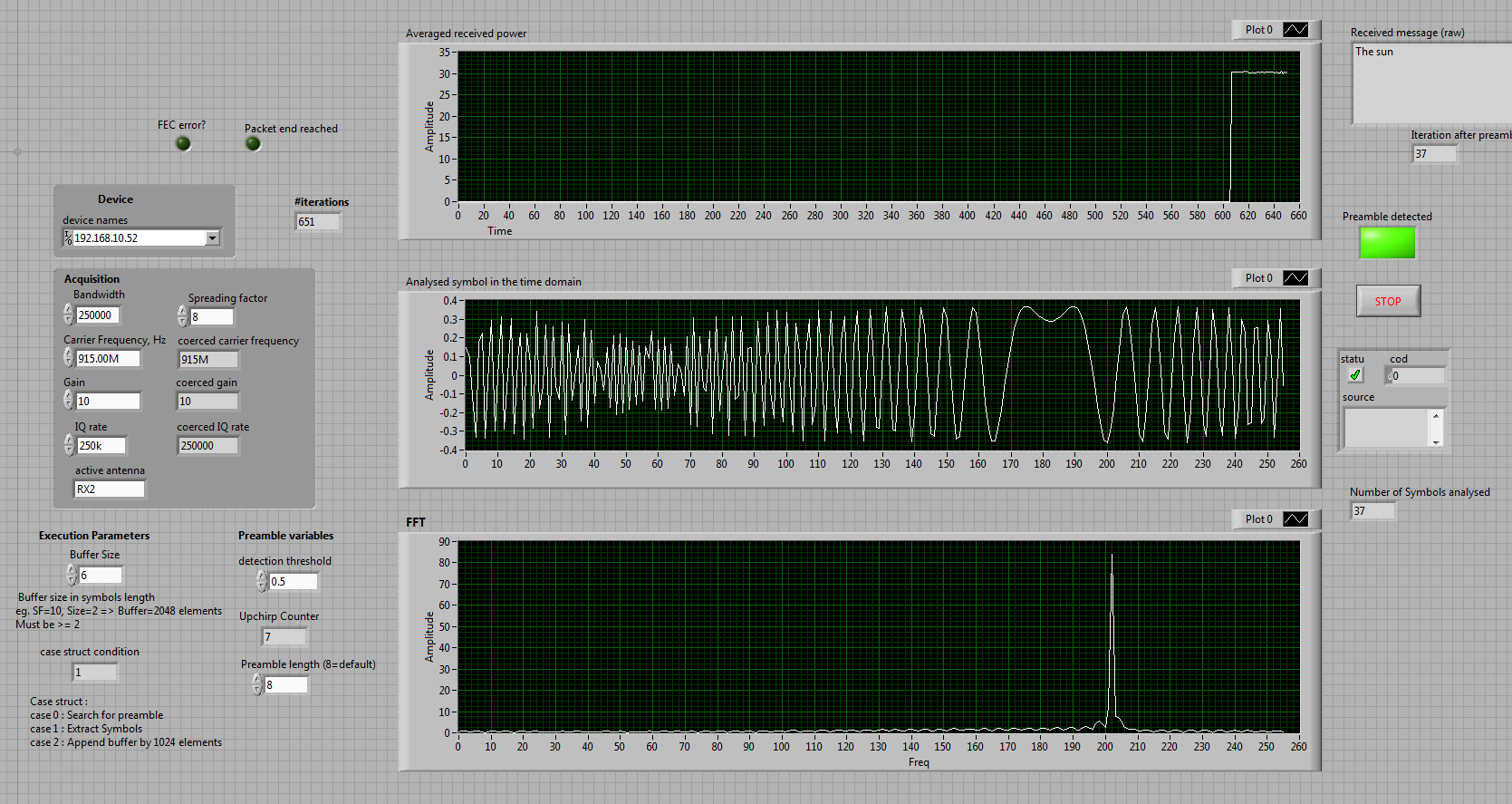}}
	\vspace{-3mm}
	\caption{Implementation set-up and the received LoRa signal.}
	\label{fig:GUI}
	\vspace{-3mm}
\end{figure}

\rdbsec
\rdextra
\section{Conclusion}\label{sec:conc}

In this paper, we have provided an in-depth analysis of the LoRa PHY by analytically studying the algorithmic aspects of a LoRa transceiver.
We showed that the LoRa receiver is robust against the \gls{CFO} by synchronizing with a time-offset to the preamble, while a residual offset needs to be compensated.
Further, the \gls{SFO} effect on the LoRa demodulation was modeled as an increasing phase mismatch,
which can be prevented by re-aligning the symbol boundaries.
Finally, we showed that LoRa can be implemented and tested on a USRP platform. 

\vfill\pagebreak

\bibliographystyle{IEEEtran}
\bibliography{./share/IoTJabref}

\end{document}